\documentclass[twocolumn,showpacs,preprintnumbers,amsmath,amssymb,floatfix]{revtex4}
\usepackage{graphicx}
\usepackage{dcolumn}
\usepackage{bm}
 
\usepackage{epsfig} 
\usepackage {array} 
\usepackage {amsmath} 
\usepackage {lscape}
\providecommand\bb{\bm{{\rm b}}} 
\providecommand\bB{\bm{{\rm B}}}

\providecommand\bj{\bm{{\rm j}}}

\providecommand\bU{\bm{{\rm U}}}

\providecommand\rm{R_m} 
\begin{document} 
\preprint{APS/123-QED}
\title{Influence of electro-magnetic boundary conditions onto \\
the onset of dynamo action in laboratory experiments}
\author{Raul Avalos-Zuniga}
\author{Franck Plunian}
\email{Franck.Plunian@hmg.inpg.fr}
\homepage{http://legi.hmg.inpg.fr/~plunian}
\affiliation{
Laboratoires des Ecoulements G\'{e}ophysiques et
Industriels,B.P. 53, 38041 Grenoble Cedex 9, France \\
}
\author{Agris Gailitis}
\affiliation{
Institute of Physics,University of Latvia,LV-2169 Salaspils 1,Riga district,Latvia
}
\begin{abstract} 
We study the onset of dynamo action of the Riga and Karlsruhe
experiments with the addition of an external wall, the
electro-magnetic properties of which being different from those of the fluid in
motion. We consider a wall of different thickness, conductivity and
permeability. We also consider the case of a ferro-fluid in motion.
\end{abstract}
\maketitle
\section{Introduction}
\subsection{Objectives}
Two dynamo experiments have been successful so far, one in Riga (Latvia)
\cite {Gai00,Gai01} and one in Karlsruhe (Germany) \cite {Sti01}.
Both experiments are complementary to each other in many
respects. One is mono-cellular with a dynamo mechanism based on a laminar
kinematic approach. The second is multicellular with scale separation between
the flow and the magnetic field leading to an alpha-effect as assumed in turbulent
dynamos. The first one produces a time-dependent magnetic field (Hopf
bifurcation) whereas the second one produces a stationary magnetic field
(stationary bifurcation). Finally in both cases the theoretical predictions
proved to be in very good agreement with the experimental results. This gives
good confidence for further theoretical investigations as it is done in this
paper.\\ We address questions about the
influence of electro-magnetic boundary conditions onto the onset of dynamo
action. Suppose for example that an external layer of stagnant fluid is added
around the main motion as it is done in Riga. Does it help for dynamo action?
What does happen if instead of stagnant fluid the external layer is a highly
conducting wall or a ferromagnetic wall (with a magnetic permeability larger
than vacuum permeability)? At last what is the influence onto the onset of
dynamo action when a ferro-fluid is used (assuming a homogeneous permeability
in all the fluid) as proposed recently \cite {Fri02}?\\
The answers to these questions are of high interest for the next generation of
dynamo experiments which are in preparation \cite {Gai02,Ceb02}. Indeed, with
concern for natural dynamos, these new generation experiments do not have a
flow geometry as well optimized as the two previous ones. Then the volume of
moving liquid metal necessary to get dynamo action is much larger. In fact this
volume may even be underestimated by the theoretical predictions usually based
on crude approximations as laminarity of the flow. Then the possibility to add
external walls or stagnant fluid around the experiment 
as well as the use of a ferro-fluid could become essential. 
\subsection{Geometries of Riga and Karlsruhe experiments}
\begin{figure}   
  \begin{tabular}{@{\hspace{0em}}c@{\hspace{-2.5em}}c@{}}    
    \epsfig{file=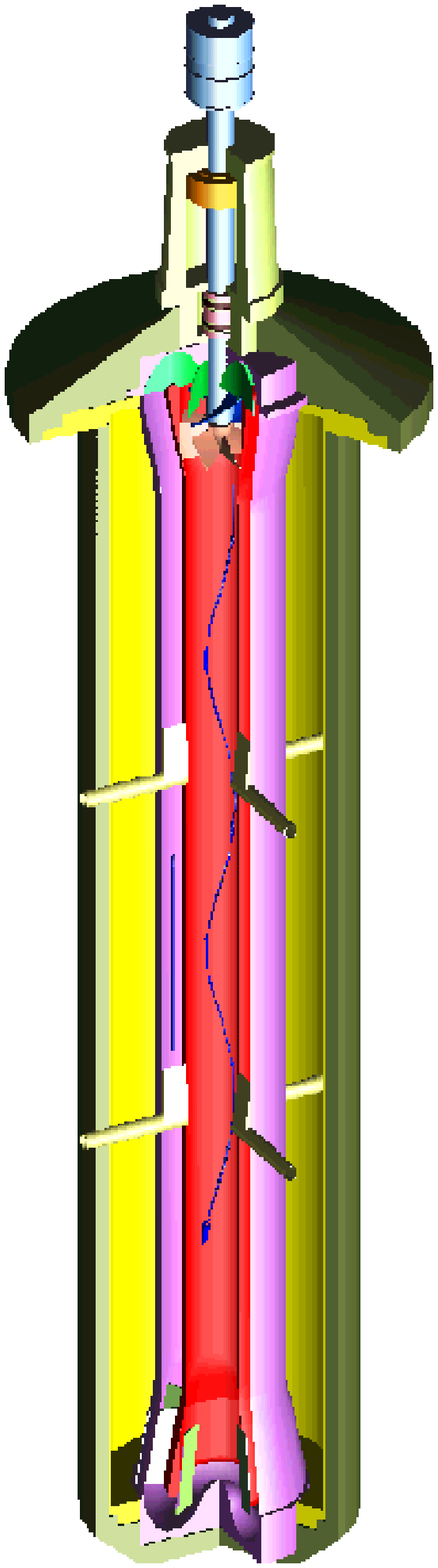,width=0.1\textwidth}&    
    \epsfig{file=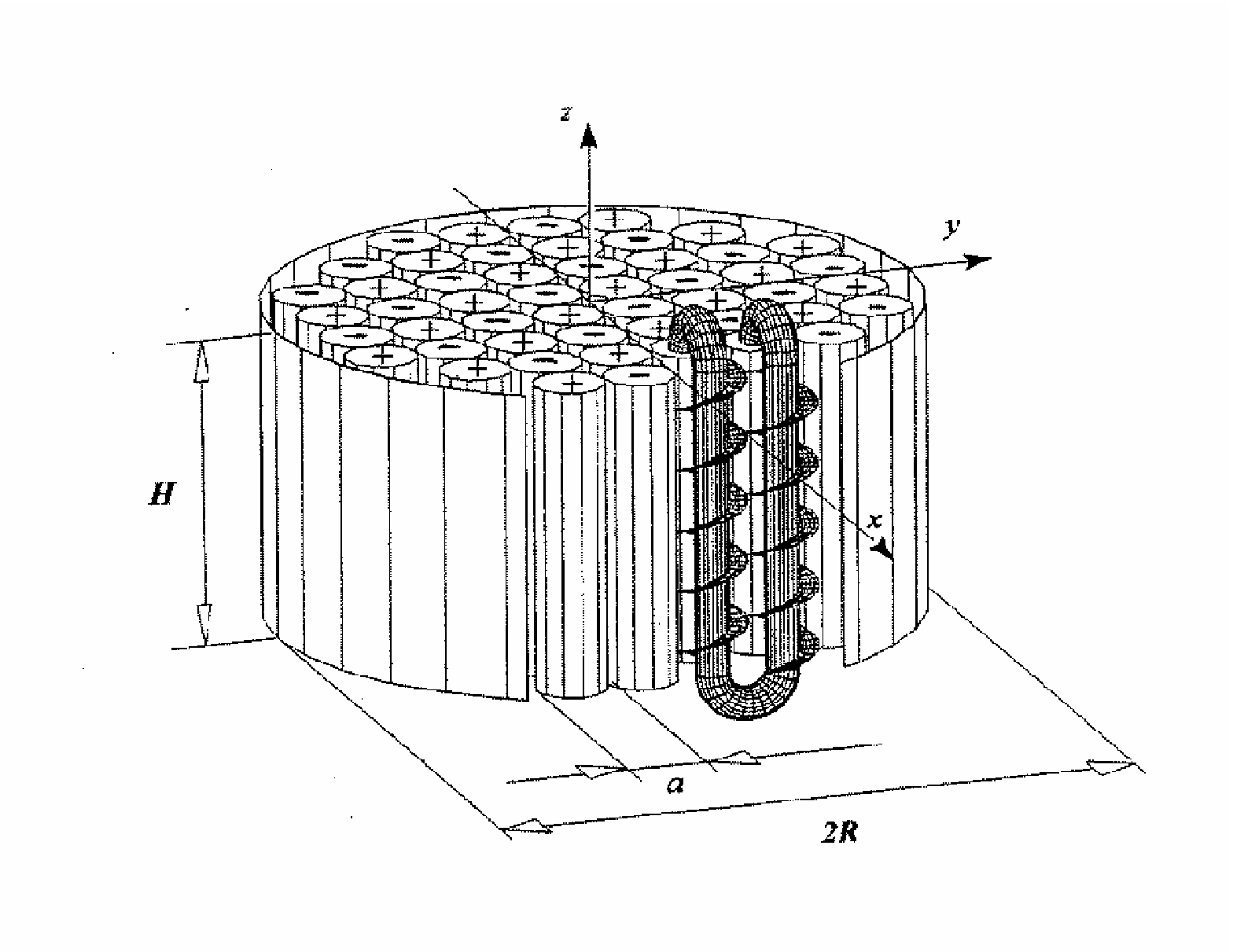,width=0.38\textwidth}\\
    (a)&\parbox{0.48\textwidth}{\makebox[4.5cm]{(b)}}
    \end{tabular}
\caption{The dynamo modules of the (a)
Riga and (b) Karlsruhe experiments.} 
\label{fig:exp}
\end{figure}

For both experiments the appropriate coordinates are cylindrical
($r,\theta,z$).\\
The Riga dynamo experiment \cite{Gai00} is composed
of three co-axial cylinders of radius $r_0=0.125$m, $R=0.215$m and $R+e=0.4$m.
The flow is helical in the inner cylinder, straight and backwards between the
inner and the second cylinder (Fig.\ref{fig:exp}a). There is stagnant fluid
in the outer cylinder. The same fluid (liquid sodium) has been used in the
different parts of the experiment.   The height of the device is $H=2.91$m.\\
The essential piece of the Karlsruhe dynamo experiment
\cite{Sti01} is a cylindrical
container with both radius $R$ and height $H$ somewhat less than 1 m, through
which liquid sodium is driven by external pumps. By means of a system of
channels, constituting 52 "spin generators",  a helical
motion is organized (Fig.\ref{fig:exp}b). The flow pattern is of Roberts
\cite{Rob72} type and an estimate of the self-excitation condition for this
experimental device has been derived from a mean-field solution with an
$\alpha$-effect assumed to be constant in the cylinder \cite{Rad01}.
\\
\section{Formulation of the problem}
\subsection{Parameters}
For our calculations we consider three co-axial cylindrical regions
defined by their radius ($r_1=R, r_2=R+e, r_3=\infty$), their
conductivity ($\sigma_1,\sigma_2,\sigma_3$) and permeability ($\mu_1, \mu_2,
\mu_3$). The region 1 contains
the moving fluid, the region 2 is the conducting wall (or stagnant
surrounding fluid) and the region 3 is the insulator around the experiment
($\sigma_3=0$). However, for sake of generality we will replace $\sigma_3$ by
zero only in the numerical applications. 
\subsection{Kinematic dynamo problem}
As we are interested in the onset of the dynamo
instability, it is sufficient to solve the kinematic dynamo problem in which
the flow is considered as given.
The magnetic field
$\bB$ must satisfy the induction equation and the divergence-free condition
\begin{eqnarray}
\frac{\partial \bB}{\partial t} = \nabla \times (\bU \times \bB) - \nabla
\times ([\alpha] \bB)+ (\mu \sigma)^{-1}\nabla^2 \bB,\label{eq:ind}\\
\nabla \cdot \bB = 0, 
\end{eqnarray}
with appropriate boundary conditions (as we shall see later)
and where the velocity
field $\bU$ and the [$\alpha$]-tensor may be non zero only in the region 1. The
[$\alpha$]-tensor corresponds to a mean electromotive force which is linear
and homogeneous in $\bB$. In that case the quantities $\bB$ and $\bU$ must be
understood as mean quantities \cite{Kra80}.
\subsection{Velocity and [$\alpha$]-tensor}
In the Riga experiment the velocity $\bU$ is defined by
$\bU_0=(0,\omega r,\chi \omega r_0)$ for $r\le r_0$ and $\bU_1=(0,0,-\frac{\chi
\omega r_0}{(R/r_0)^2-1})$ for $r_0<r\le r_1$. Therefore it is convenient to introduce an additional
cylindrical region 0 defined by its radius $r=r_0$ distinct from region 1
($r_0<r\le r_1$) by its velocity but common by its conductivity
$\sigma_0=\sigma_1$ and permeability $\mu_0=\mu_1$ (as it is the same fluid).
The [$\alpha$]-tensor is identically zero at first order for the Riga
experiment. Indeed the currents induced by the small scale of the turbulence
are negligible compared to the currents induced by the mean flow.\\
For the
Karlsruhe experiment, it is the mean flow $\bU$ which is zero. In that case,
the [$\alpha$]-tensor writes $\alpha_{ij}=\alpha_{\perp}(\delta_{ij}
-e_ie_j)$. This corresponds to an anisotropic $\alpha$-effect deduced from the
symmetry properties of the flow. In addition in the calculation of the mean
electromotive force we neglected the contribution which contains the derivatives of $\bB$. This
approximation leads to an error of about $10\%$ on the instability threshold
prediction \cite{Rad01,Plu02}. However this approximation is accurate enough
for our present purpose.\\ 
For convenience we denote each region by $l$
(=1, 2 or 3 plus the additional region $l$=0 for the Riga experiment).
\subsection{Magnetic field}
As
the flow in both problems is $z$-independent, axisymmetric and
time-independent, a particular solution of  (\ref{eq:ind}) takes the form  
\begin{equation}  
\hat{\bB}(r,\theta,z,t)=\bb(r)\;e^{pt+im\theta
+ikz},  
\label{eq:bb}  
\end{equation}
$p$ being the complex growth rate, $m$ and
$k$ the azimuthal and vertical wave numbers.  
The superposition of all the
$(m,k)$-modes $\hat{\bB}$ leads to the general solution $\bB$ of (\ref{eq:ind}) 
to which the boundary conditions apply.\\
The radial boundary conditions write 
$\lim_{r\rightarrow \infty}\bb_3=0$ plus the
appropriate relations between each region $l$ (see below). 
As these relations are satisfied by each particular solution 
$\hat{\bB}$ they are also satisfied by
$\bB$.\\
The axial boundary
conditions write  
\begin{eqnarray}
\lim_{z\rightarrow \pm\infty}\bB=0 \label{bcinf}\\
(\nabla\times\bB)_z=0\;\; \mbox{at}\;\; z=\pm H/2,\label{bcjz}
\end{eqnarray}
(\ref{bcjz}) meaning
that there is no axial current crossing the  
insulating borders at both ends.
In order to simplify the calculations we
shall consider only two $(m,k)$-modes $\hat{\bB}$,
the superposition of which satisfying
(\ref{bcjz}) only, as explained
later in the paper.  
\section{Method of solution}
\subsection{Solutions of the dynamo problem}
When replacing (\ref{eq:bb}) in (\ref{eq:ind}) we find
that in each region $l$ the radial and azimuthal components of $\bb$ must
satisfy 
\begin{equation}
\bb_{l}''+\frac{1}{r}\bb_{l}'+[A]\bb_{l}=0
\label{eq:inducbl}
\end{equation}
with the prime denoting the $r$-derivative and
where the matrix $A$ simply relates the different components of $\bb_{l}$ in
the induction equation. The $A$ coefficients write
$A_{11}=A_{22}=-(k^2+\frac{p}{\eta_l}+\frac{m^2+1}{r^2}+i\frac{m\omega_l+kV_l}
{\eta_l})$ and $A_{12}=-A_{21}=-i(\frac{2m}{r^2}+\frac{k\alpha_l}{\eta_l})$,
with $\eta_l=(\sigma_l\mu_l)^{-1}$ and where 
$\alpha_l, \omega_l$ and $V_l$ are
the magnetic diffusivity, the $\alpha$-effect, the rotation rate and the
$z$-component of the velocity field appropriate to each region $l$ and to each
case (Riga or Karlsruhe) as defined above. Finally the component $b_{lz}$ can
be determined subsequently by: 
\begin{equation}
b_{lz}=\frac{i}{k}(\frac{b_{lr}+imb_{l\theta}}{r}+b_{lr}'). \label{blz}
\end{equation}
To find the solutions in the
region $l$, instead of ($b_{lr},b_{l\theta}$) we look for ($b_{lr}+i
b_{l\theta},b_{lr}-ib_{l\theta}$). These solutions can be written as a linear
combination of modified Bessel's functions $I_{m+1}(\omega_l^+ r)$ and
$K_{m+1}(\omega_l^+ r)$ for $b_{lr}+i b_{l\theta}$ and $I_{m-1}(\omega_l^- r)$
and $K_{m-1}(\omega_l^- r)$ for $b_{lr}-i b_{l\theta}$, where 
\begin{equation}  
(\omega_l^\pm)^2=k^2+\frac{p\pm\alpha_lk+i(m\omega_l+kV_l)}{\eta_l}.
\label{omegal}
\end{equation} 
In each region the solutions write in the form
\begin{eqnarray} 
(b_{lr},ib_{l\theta})=(F^+_lI_{m+1}(\omega_l^+
r)+G^+_lK_{m+1}(\omega_l^+r))(1,1)\nonumber \\                     
+(F^-_lI_{m-1}(\omega_l^- r)+G^-_lK_{m-1}(\omega_l^-r))(1,-1) 
\label{brbteta} 
\end{eqnarray}
where $F^+_l$, $F^-_l$, $G^+_l$ and $G^-_l$ are constants.
The regularity conditions for $r=0$ lead to 
$G^+_0=G^-_0=0$ for the Riga experiment and $G^+_1=G^-_1=0$ for the Karlsruhe
experiment.
The condition $\bb_3 \rightarrow 0$ when $r
\rightarrow \infty$ leads to $F^+_3=F^-_3=0$ for both experiments.
\subsection{Radial boundary conditions}
The normal component of $\bB$, the tangential component of $\bB/\mu$ and the
$z$-component of the electric field $E_z=\eta (\nabla \times \bB)_z$ are
continuous across each interface $r=r_0$ (only for Riga), $r=r_1$ and $r=r_2$.
We can show that this set of relations is sufficient to describe all the radial
boundary conditions of the problem. They write at $r=r_l$:  
\begin{eqnarray}  
b_{l,r}&=&b_{l+1,r}\nonumber \\    
\frac{b_{l,\theta}}{\mu_l}&=&\frac{b_{l+1,\theta}}{\mu_{l+1}}\nonumber\\
\frac{1}{\mu_l}(\frac{b_{l,r}}{r_l}+b_{l,r}')&=&\frac{1}{\mu_{l+1}}(\frac{b_{l+1,r}}{r_l}+
b_{l+1,r}')\nonumber\\
 \eta_l(\frac{b_{l,\theta}-im
b_{l,r}}{r_l}+b_{l,\theta}')&=&\eta_{l+1}(\frac{b_{l+1,\theta}-im
b_{l+1,r}}{r_l}+b_{l+1,\theta}')\nonumber\\
\label{BC}
\end{eqnarray}
but for the Riga experiment at $r=r_0$ the last equation in (\ref{BC}) is
replaced by:
\begin{eqnarray}  
 (b_{1\theta}')&=&(b_{0\theta}'+\frac{1}{\eta_1}\omega
r_0b_{0r}). \label{BC2}
\end{eqnarray}
\subsection{Dispersion relation and dimensionless parameters}
Replacing (\ref{brbteta}) into (\ref{BC}) and (\ref{BC2}), we
find a system of eight equations for the Karlsruhe dynamo and twelve for the
Riga dynamo. We have a non trivial solution
only if the determinant of the system is equal to zero. This writes in the form: 
\begin{equation} 
F(R_m (\mbox{or} R_\alpha), k, p, m, \mbox{geometric 
parameters})=0 
\label{eq:F} 
\end{equation}  
where $R_m$ and $R_\alpha$ are 
magnetic Reynolds numbers defined by 
$R_m=\sigma_1\mu_1 |\bU_0|_{max} r_0$ 
for 
the Riga dynamo and $R_\alpha=\sigma_1\mu_1 \alpha_{\perp}R$ for the 
Karlsruhe 
dynamo. \\
For the 
calculations we set $\sigma_3=0$ and define $\sigma_2/\sigma_1=s, 
\mu_1/\mu_3=q$ and  $\mu_2/\mu_3=n$. The dynamo onset corresponds to
$\Re(p)=0$ for  which a critical $R_m$ or $R_\alpha$ is calculated for
different values of  the parameters $e/R$, $s$, $q$, $n$ and for values of $k$
chosen to satisfy the axial boundary condition (\ref{bcjz}) as explained
below.  Like any  transcendental 
equation (Bessel functions with complex arguments), (\ref{eq:F}) has an 
infinite number of complex roots.
It has to be solved numerically. \\
\subsection{Treatment of the axial boundary condition}
\subsubsection{Method}
Any  $(m,k)$-mode $\hat{\bB}$ satisfying (\ref{eq:F})
automatically satisfies the radial boundary conditions but not the
axial boundary condition. For that, again,
one should write
$\bB$ as the superposition of an infinite number of particular solutions
$\hat{\bB}$ satisfying (\ref{eq:F}) and then apply (\ref{bcinf}) and
(\ref{bcjz}) to $\bB$.   This is quite tedious and numerically demanding.
Instead we look for an approximate solution $\bB$ written as the superposition
of only two particular solutions $\hat{\bB}_1$ and $\hat{\bB}_2$ which have
the same growth rate $p$ and with wave numbers $k_1$ and $k_2$ which difference
 writes \begin{equation}k_1-k_2=2\pi/H.  \label{eq:D} \end{equation} 
If in addition both solutions have the same radial profile then 
(\ref{bcjz}) is satisfied, which is a good enough
approximation of the actual experiments. 
Such an approximation is quite well justified for the Riga experiment 
in reason of its extended  shape $H/R \sim 15$. Indeed, as 
the radial profile difference between both solutions at
$z=\pm H/2$ is of the order $O(R/H)$,
the boundary conditions (\ref{BC}) and (\ref{BC2}) are satisfied with 
an error also of $O(R/H)$ and the parameters in (\ref{eq:F}) 
are obtained with an error of the order
$O(R^2/H^2)$. 
In the case of Karlsruhe ($H/R \sim 1$), the only justification is common
experience that  in  many similar cases replacing zero boundary conditions at
infinity by periodic boundary conditions at both ends (leading to
(\ref{eq:F})) makes no crucial difference.
\subsubsection{Karlsruhe}
With such an
approximation the problem is straightforward to solve for the Karlsruhe 
experiment.  Indeed as the flow pattern is symmetric to the plane $z=0$, after 
(\ref{eq:F}) the two solutions with $k=\pm \pi/H$ have the same $p$ and 
satisfy  (\ref{eq:D}). The clockwise and anticlockwise  rotations are
compensated implying that the generated field pattern does not  rotate round
the symmetry axis. Hence the equation (\ref{eq:F}) can be  written in real
variables, the growth rate $p$ is real and the field  is a stationary field.  An
other way to understand it is that the $\alpha$-effect  does  not depend on
$z$ and therefore there is no preferred sense in the  $z$-direction 
for a magnetic wave to travel as it would be if not stationary. Then the only 
thing which 
remains to do is solving (\ref{eq:F}) in order to find the critical $R_\alpha$
for which $p=0$. For  the 
calculation we took $H/R=1$.
\subsubsection{Riga}
For the Riga experiment the calculation is more complicated than for Karlsruhe
for at least two reasons.\\
First, the inner flow ($r < r_0$) is helical and has then a preferred
direction  given by the rotation axis.
Then any generated field pattern rotates
round the  vertical axis of symmetry. Hence the field is not stationary and the
growth rate  $p$ is always complex. \\
Second, one does not obtain the same result when $V_z$ is replaced by $-V_z$ in
both regions 0 and 1. This implies that $p(-k)$ is always different
from $p(k)$ contrary to the Karlsruhe case.\\
As a result, for a given $R_m$ one
must look for two complex values of $k$ which only differ from their real
parts while their  imaginary parts are equal (see \cite{Gai90} for more
details) and which must satisfy (\ref{eq:F}),
(\ref{eq:D}) and $p(k_1,R_m)=p(k_2,R_m)$. The generated instability is usually
known as absolute or global instability. 
The generated magnetic field $\bB=\hat{\bB}_1+\hat{\bB}_2$ is a deformed (as
$\Im(k_1)=\Im(k_2)\ne 0$) standing wave damped at both ends of the device and
rotating around the symmetry axis. 
We call absolute critical $R_m$
the value of $R_m$ such that 
these conditions plus the additional relation
$\Re(p)(k_1,R_m)=\Re(p)(k_2,R_m)=0$ are satisfied.
At the time when the Riga experiment was designed,
this method had already been used. In particular
the size $r_1-r_0$ of the Riga experiment
was determined to lower the group velocity $v_G=i\partial p/\partial k$ 
of the above mentioned absolute instability.
For our calculations we used the values of $r_0$,$r_1(=R)$ and $H$
as given above and $\chi=1$ which is representative of the actual
flow of the Riga experiment \cite{Gai00}. 
\section{Results}
\subsection{Integral quantities}
In all our calculations for both Riga and Karlsruhe the azimuthal mode
$m=1$ has always been found to be dominant. Therefore in the rest of the
paper only the results for this mode are presented.
From now $R_m$ (resp. $R_{\alpha}$)
denotes the absolute critical $R_m$ (resp. critical $R_{\alpha}$).\\ In order
to give some physical justification of our results we need to define the
additional following quantities ${\cal W}_l$, ${\cal P}_l$, ${\cal J}_l$ and
${\cal S}_l$ which are respectively the magnetic energy, the Poynting flux,
the Joule dissipation and the work of the Lorenz forces in the region
$\Omega_l$ ($l=1$ for the fluid, $l=2$ for the wall and $l=3$ for the vacuum).
They are defined by:   \begin{eqnarray}
{\cal W}_l=\int_{(\Omega_l)}\frac{B^2}{2\mu}d\Omega&,&
{\cal P}_l=\int_{(S_l)}(\frac{\bold B}{\mu}\times {\bold E})\cdot {\bold
n}dS,\\ {\cal J}_l=\int_{(\Omega_l)}\frac{j^2}{\sigma}d\Omega&,&
{\cal S}_l=\int_{(\Omega_l)}{\bold j}\cdot {\cal E}d\Omega
\end{eqnarray} 
where ${\cal E}={\bold U}\times {\bold B}$ for Riga and 
${\cal E}=-[\alpha]{\bold B}$ for Karlsruhe. The region $(\Omega_l)$ is
delimited by the boundary(ies) $(S_l)$ of normal $\bold n$ and $\bold
j=\nabla\times \bB/\mu$ is the current density. Multiplying (\ref{eq:ind}) by
${\bold B}/\mu$ and integrating in each region $l$ we find:
\begin{eqnarray}
\frac{\partial {\cal W}_1}{\partial t}= {\cal P}_1+{\cal S}_1-{\cal J}_1 &,&
\frac{\partial {\cal W}_2}{\partial t}= {\cal P}_2-{\cal J}_2\\
\frac{\partial {\cal W}_3}{\partial t}= {\cal P}_3&,& {\cal P}_1+{\cal
P}_2+{\cal P}_3=0.
\end{eqnarray} 
Dynamo action corresponds to 
\begin{equation}
{\cal S}_1 \ge{\cal J}_1+{\cal J}_2
\label{SJ}
\end{equation}
with the
equal sign for the instability threshold . 
It means that at the threshold the work of the Lorenz forces ${\cal S}_1$
must compensate the total ohmic dissipation.
\subsection{Rigid body helical flow}
Before dealing with the Riga and Karlsruhe experiments we first
want to mention results for the academic case of a rigid body helical flow
surrounded by a conducting wall, both having infinite height. This case
corresponds to  have  $r_0=R$ in our calculations for the Riga geometry (in
that case the region 1  of  the backwards flow does not exist). However
instead of looking for an absolute instability as for Riga we simply look for
the onset of the dynamo instability corresponding to the minimum value of
$R_m$ for a given $k$.  This instability is found to be convective,
any primordial perturbation when growing being also traveling along the
vertical  axis of symmetry. \\
We repeated the
results of \cite{Gai80} on the dependence on conductivity and thickness. 
A decrease of the dynamo  threshold has been found as the
dimensionless wall thickness $e/R$ or wall  conductivity $s$ was increased. The
usual picture to  explain this result is that increasing the wall thickness or
wall  conductivity  leads in both cases to a reduction of the ohmic
dissipation.   From (\ref{SJ}) the reduction of the total
dissipation ${\cal J}_1 + {\cal J}_2$ is equivalent to the reduction of ${\cal
S}_1 $ which is directly related to the threshold.\\
In the case of uniform 
conductivity $s=1$, it has been shown \cite{Kai99} that this picture is 
incomplete  when the magnetic field is time-dependent. In that case some
additional eddy currents may be induced in the wall, increasing the ohmic
dissipation. As a result the dynamo threshold versus the wall thickness has a
minimum.\\ 
In our calculations we checked out the existence of this minimum. We found that
this  effect is even more important for $s>1$. We found a similar effect for
Riga as explained in the next section.
  \subsection{Influence of the wall conductivity} \subsubsection{Threshold
reduction rate} 
\begin{figure}   
\begin{flushright}   
\begin{tabular}{@{\hspace{0cm}}l@{\hspace{0cm}}c@{\hspace{0cm}}}     
\raisebox{4cm}{$\Gamma$}    \epsfig{file=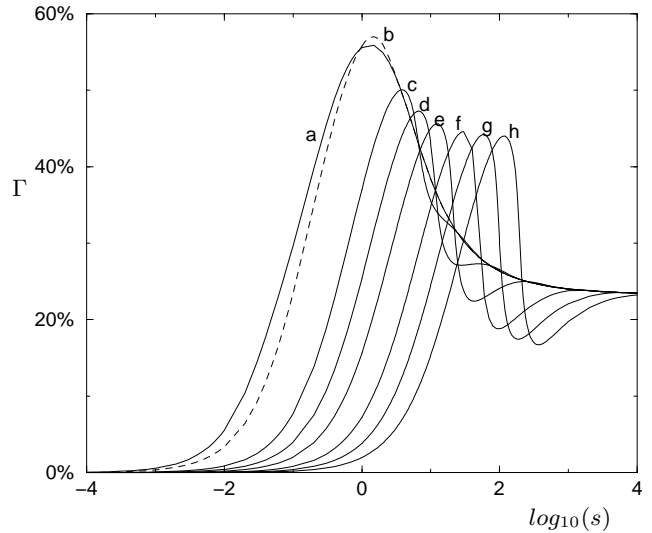,width=0.45\textwidth} 
    \\*[0cm] 
     & \parbox{0.5\textwidth}{\hspace{-11cm} $log_{10}(s)$} 
    \end{tabular} 
\end{flushright} 
    \caption{Riga: The threshold reduction rate $\Gamma$
versus $log_{10}(s)$ for $n=q=1$ and different values of $e/R$.
For curve (a) the ratio $e/R$ is infinite, for (b) 86 \% (dashed line), for (c) 20 \%, for
(d) 10\%, for (e) 5\%, for (f) 2 \%, for (g) 1\% and for (h) 0.5 \%.} 
\label{fig:rigasigma}    
\end{figure}

\begin{figure} 
  \begin{flushright} 
  \begin{tabular}{@{\hspace{0cm}}l@{\hspace{0cm}}c@{\hspace{0cm}}} 
    \raisebox{5cm}{$\Gamma$}
  \epsfig{file=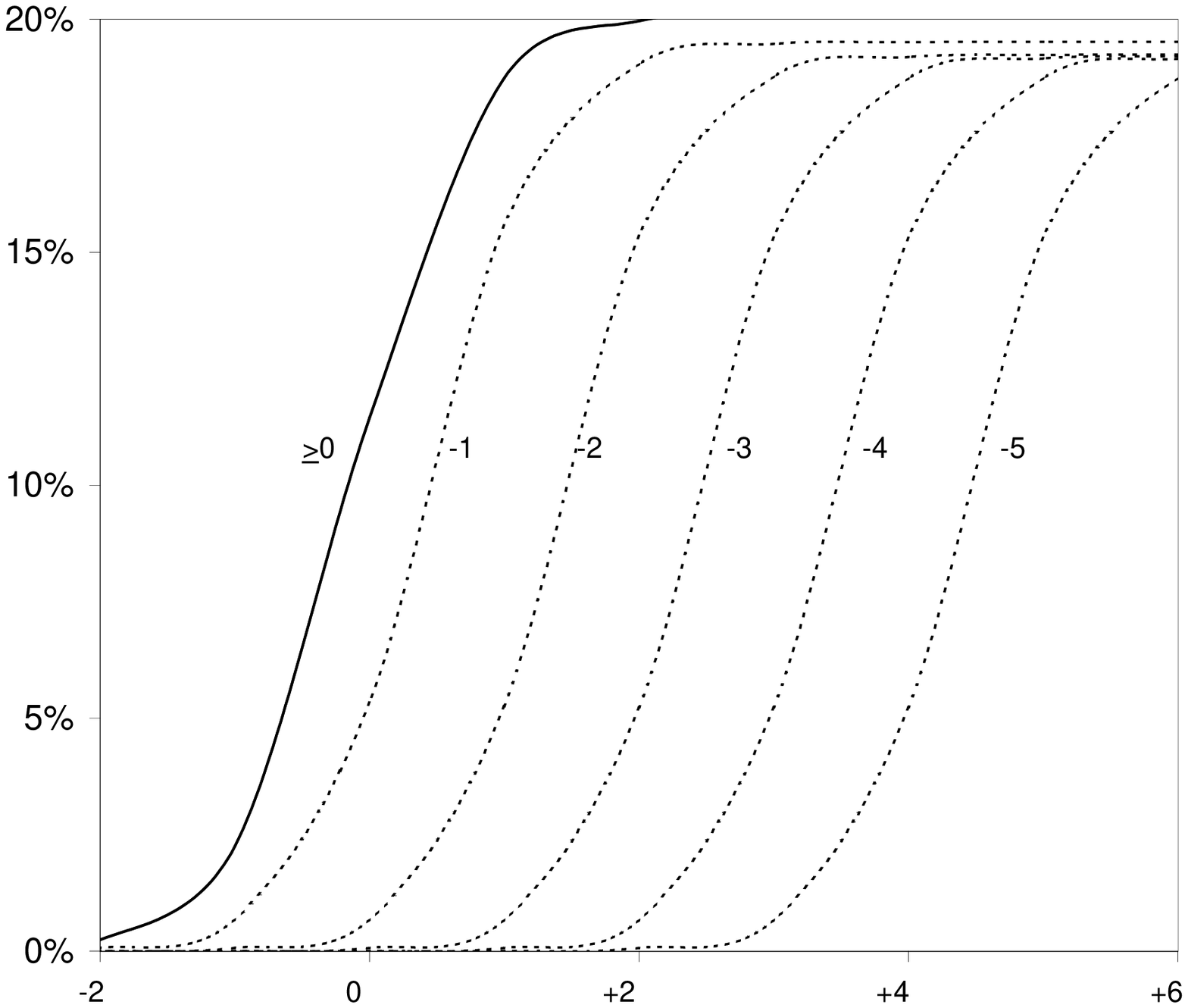,width=0.7\textwidth}
    \\*[-1cm] 
     & \parbox{0.5\textwidth}{\hspace{-23cm} $log_{10}(s)$}
      \end{tabular} 
\end{flushright}
    \caption{Karlsruhe: The threshold reduction rate $\Gamma$
versus $log_{10}(s)$
for $n=q=1$. The labels indicate $log_{10}(e/R)$.}  
\label{fig:karl}      
\end{figure}

To present our results we adopt the
point of view of any experimenter who wants to know how much reduction of the
dynamo threshold he can obtain varying the wall thickness and conductivity,
relatively to the case with no wall at all ($e=0$). For that we
define a threshold reduction rate by 
\begin{equation}
\Gamma=1-\frac{R_m(s,e/R)}{R_m(e=0)} \label{gain}
\end{equation}
for Riga which also applies to Karlsruhe replacing $R_m$ by $R_{\alpha}$.
We found $R_m(e=0)=41.16$ for Riga and $R_{\alpha}(e=0)=4.8$ for Karlsruhe. 
The reduction rates for Riga and Karlsruhe are plotted
respectively in Fig.\ref{fig:rigasigma} and  Fig.\ref{fig:karl} versus $s$ for
$n=q=1$ and for different wall dimensionless thicknesses $e/R$.\\ 

In both cases $\Gamma$ is always positive which stresses the interest of
having a conducting wall. Of course $\lim_{s \rightarrow 0} \Gamma = 0$.
Indeed, as the wall is surrounded by the vacuum, having a non conducting wall
is equivalent to have no wall at all. In both cases
$\lim_{s \rightarrow \infty} \Gamma \approx 20 \%$.
\subsubsection{The particular case $s=1$ for Riga}
For $log_{10}(s)=0$, the maximum reduction rate obtained for Riga is
$55.9\%$. This is surprisingly close to the value obtained for a spherical
dynamo model surrounded by a quiescent conducting external shell considered in
\cite{Sar96}. In table 8 of \cite{Sar96}, they found $R_m(e=0)=3901.11$ and
$R_m(e=\infty)=1659.05$ leading to $\Gamma=57.5\%$.\\
A remarkable point for Riga is that the choice adopted for
the experiment, $e/R=86\%$ (curve b) and $s=1$ leads
to the maximum threshold reduction rate. This shows that 
there is no benefit of adding
a high electro-conducting wall instead of an outer stagnant layer
of fluid.\\
In Fig.\ref{fig:rigasigma} the dashed-curve (b) goes above the solid-curve (a) for
$s=O(1)$. This shows that there is a wall thickness ($\approx 86 \%$ for Riga)
for which the dissipation is minimum. For a larger thickness (curve a)
additional dissipation occurs, probably in reason of additional eddy currents
as found in \cite{Kai99} for the rigid body helical flow. We recall here
that this effect is related to the time-dependency of the
solution. This would explain why such a curves crossing is observed for Riga
(time-dependent solution) and not for Karlsruhe (stationary solution).  
\subsubsection{Physical interpretation} 
In this section we give some physical interpretation on the behavior of
$\Gamma$ versus $s$. In a first step let us consider the case of Karlsruhe for
which $\Gamma$ increases monotonically with $s$. From (\ref{SJ}) the threshold
is directly related to ${\cal J}_1$ and ${\cal J}_2$ the dissipation in the
fluid and the wall.
We first show that in both cases $s<<1$ or $s>>1$ we have ${\cal J}_2 <<
{\cal J}_1$.\\
For $s<<1$ the electric currents
circulate mainly in the fluid. 
At the fluid-wall boundary we have $j_1 \approx j_{t1}$ and $j_2 \approx
j_{t2}$ where the subscript $t$ denotes the tangential component. Writing the
continuity of the tangential component of the electric field across the
fluid-wall boundary we find that $j_{t1} \sim j_{t2}/s$. Then integrating on
both regions (fluid and wall) we find that ${\cal J}_1 \approx
R^2j_{t1}^2/\sigma_1$  and that ${\cal J}_2 \approx Re'j_{t2}^2/\sigma_2$ with
$e'=eR/(R+e)$. Indeed when $e>>R$ it is reasonable to assume that the currents
in the wall close within a distance $R$ (instead of $e$) from the fluid-wall
boundary. As a result we find that  ${\cal J}_2/{\cal J}_1=O(se'/R)$.\\
For $s>>1$ the current lines in the fluid at the fluid-wall boundary are mainly
perpendicular to the boundary.
Therefore we have $j_1 \approx j_{n1}$ where the subscript $n$
denotes the normal component, and again $j_2 \approx
j_{t2}$ as the currents have to close up in the wall.
So we find that ${\cal J}_1 \approx R^2j_{n1}^2/\sigma_1$ 
and that ${\cal J}_2 \approx Re'j_{t2}^2/\sigma_2$.
Now from the definition of the current density $\bj=\nabla \times \bB/\mu$
we can approximate $j_{t2} \approx B_{t2}/e'\mu_2$. Writing the
continuity of $B_t/\mu$ across the
fluid-wall boundary we find that ${\cal J}_2/{\cal J}_1=O(R/se')$.\\

So we can conclude that for $se'/R<<1$ or $se'/R>>1$, the ohmic dissipation is
mainly concentrated in the fluid. Therefore from (\ref{SJ}) the threshold is
directly related to the ohmic dissipation in the fluid.  The main difference
between both limits $se'/R<<1$ and $se'/R>>1$ is the change of geometry of the
current lines in the fluid. 

\begin{figure}   
\begin{flushright}   
\begin{tabular}{@{\hspace{0cm}}l@{\hspace{0cm}}c@{\hspace{0cm}}}     
\raisebox{4cm}{$\Gamma$}    \epsfig{file=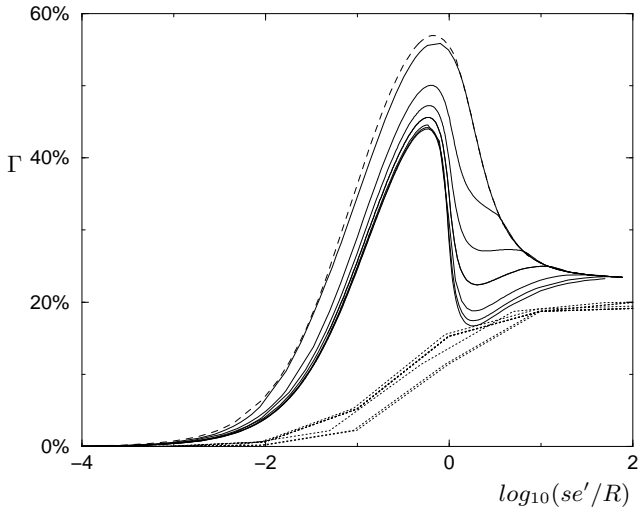,width=0.45\textwidth} 
    \\*[0cm] 
     & \parbox{0.5\textwidth}{\hspace{-11cm} $log_{10}(se'/R)$} 
    \end{tabular} 
\end{flushright} 
    \caption{Threshold reduction rate $\Gamma$
versus $log_{10}(se'/R)$ for $n=q=1$ and different values of $e/R$.
The solid (dotted) curves in the upper (lower) part correspond to Riga (Karlsruhe). The dashed curve
corresponds again to curve b of Fig.\ref{fig:rigasigma}.}  \label{fig:rigasigma2} 
   \end{figure}

For $se'/R<<1$ the current lines are constrained to close up mainly in the
fluid whereas for $se'/R>>1$ the current lines in the fluid are
perpendicular to the wall. Therefore the current lines are tighter for
$se'/R<<1$  than for $se'/R>>1$. Consequently 
we understand why the
dissipation is the largest when  $se'/R<<1$ and that $\Gamma$ increases with
$s$. Now if our argument is correct this change of geometry of the current
lines should occur at the transition between the two previous limits, namely
for  $se'/R=O(1)$. In order to check this out we
plot $\Gamma$ versus $se'/R$  in figure \ref{fig:rigasigma2}. We find that all
the curves for Karlsruhe merge pretty well (dotted curves at the bottom) and that their
change of curvature occurs indeed for $se'/R=O(1)$.\\

As a second step we consider the case of Riga for which some additional
eddy currents must be considered leading to 
an enhanced dissipation ${\cal J}_2$ concentrated in the
wall. 
Now following the same arguments than for the stationary
case, namely that as ${\cal J}_2/{\cal J}_1$ is
maximum for $se'/R=O(1)$, we expect the dissipation due to these eddy currents
to be also maximum for $se'/R=O(1)$. However in the case where the skin depth
$\delta$ is smaller than $e$, we must replace $e$ by $\delta$ in the expression
of $e'$. Indeed in case where $\delta < e$, the dissipation is mainly
concentrated in the skin layer. 
The skin depth is
defined by $\delta/R=\sqrt{2/(ns\omega)}$ where $\omega$ is the dimensionless
pulsation of the magnetic field that we also calculated solving (\ref{eq:F}).
The curves for Riga are plotted in Fig.\ref{fig:rigasigma2} (solid curves above the dotted curves
 and dashed curve at the top). 
For each thickness the maximum of $\Gamma$ is indeed obtained at about the same
value of $se'/R=O(1)$, followed by a sudden fall due to the additional
eddy currents dissipation.
Increasing the wall conductivity helps the electric currents to close outside the fluid like in the stationary case.
However because of the skin effect (non stationary solutions), increasing the wall conductivity prevents the magnetic field 
to close outside the fluid.
It is the competition between these two
effects which leads to the maximum of the threshold reduction rate $\Gamma$.\\  

\subsection{Influence of the wall permeability}
\subsubsection{Threshold reduction rate}
\begin{figure} 
  \begin{flushright} 
  \begin{tabular}{@{\hspace{0cm}}l@{\hspace{0cm}}c@{\hspace{0cm}}} 
    \raisebox{4cm}{$\Gamma$} 
  \epsfig{file=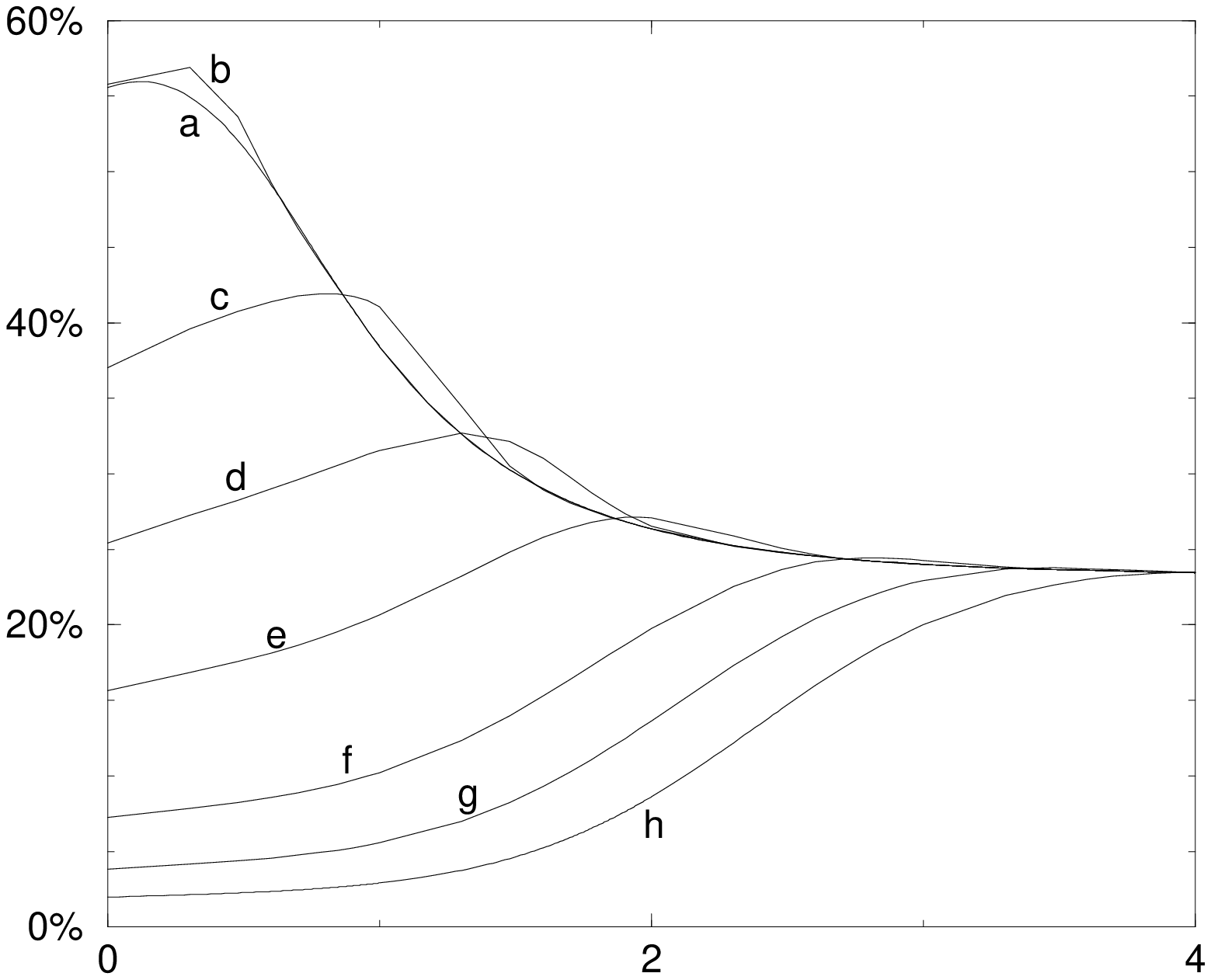,width=0.45\textwidth} 
    \\*[0cm] 
     & \parbox{0.5\textwidth}{\hspace{-11cm} $log_{10}(n)$} 
    \end{tabular} 
\end{flushright} 
    \caption{Riga: The threshold reduction rate $\Gamma$
versus $log_{10}(n)$ for $s=q=1$ and different values of
$e/R$. The labels correspond to those of Fig. \ref{fig:rigasigma}.}
\label{fig:rigamu}     
\end{figure}

\begin{figure}
  \begin{flushright} 
  \begin{tabular}{@{\hspace{0cm}}l@{\hspace{0cm}}c@{\hspace{0cm}}} 
    \raisebox{5cm}{$\Gamma$} 
  \epsfig{file=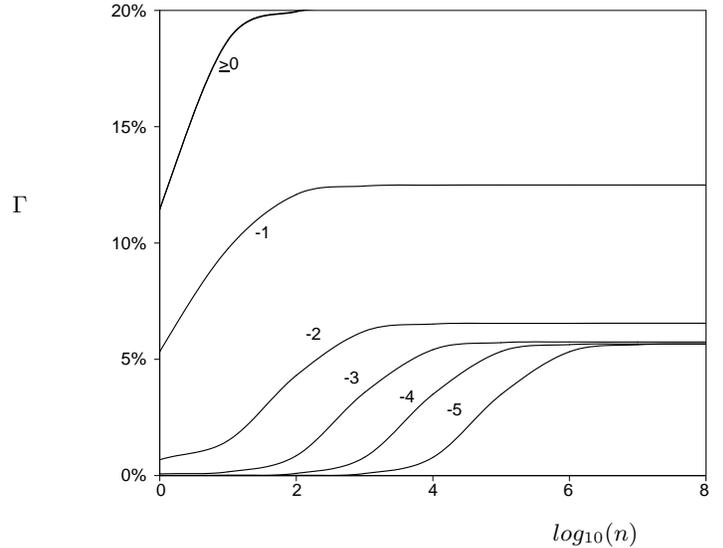,width=0.72\textwidth} 
    \\*[-1cm] 
     & \parbox{0.5\textwidth}{\hspace{-20cm} $log_{10}(n)$}
      \end{tabular} 
\end{flushright} 
    \caption{Karlsruhe: The threshold reduction rate $\Gamma$
versus $log_{10}(n)$ for $s=q=1$ and different values of
$Log_{10}(e/R)$ given by the labels.} 
\label{fig:karlmu}      
\end{figure}

\begin{figure}    \begin{flushright} 
  \begin{tabular}{@{\hspace{0cm}}l@{\hspace{0cm}}c@{\hspace{0cm}}} 
    \raisebox{4cm}{$\Gamma$} 
  \epsfig{file=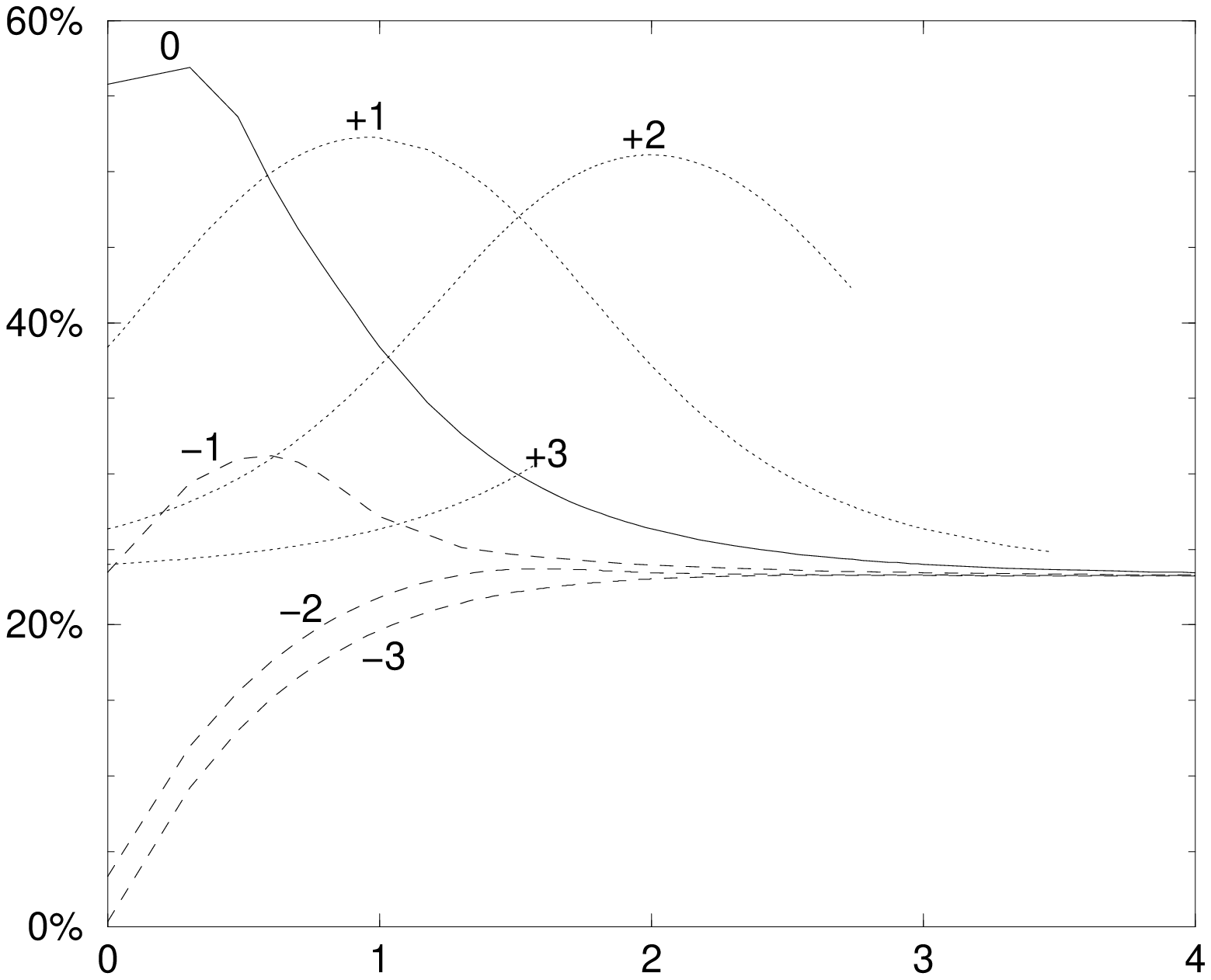,width=0.44\textwidth} 
    \\*[0cm] 
     & \parbox{0.5\textwidth}{\hspace{-11cm} $log_{10}(n)$} 
    \end{tabular} 
\end{flushright} 
    \caption{Riga: The threshold reduction rate $\Gamma$ versus $log_{10}(n)$
for $e/R=86\%,q=1$ and different values of $s$. The labels correspond to
$log_{10}(s).$ The dotted (dashed) lines refer to positive (negative)
 values of $log_{10}(s).$}
 \label{fig:rigasigmamu}     
\end{figure}

\begin{figure}
  \begin{flushright} 
  \begin{tabular}{@{\hspace{0cm}}l@{\hspace{0cm}}c@{\hspace{0cm}}} 
    \raisebox{4cm}{$\Gamma$} 
  \epsfig{file=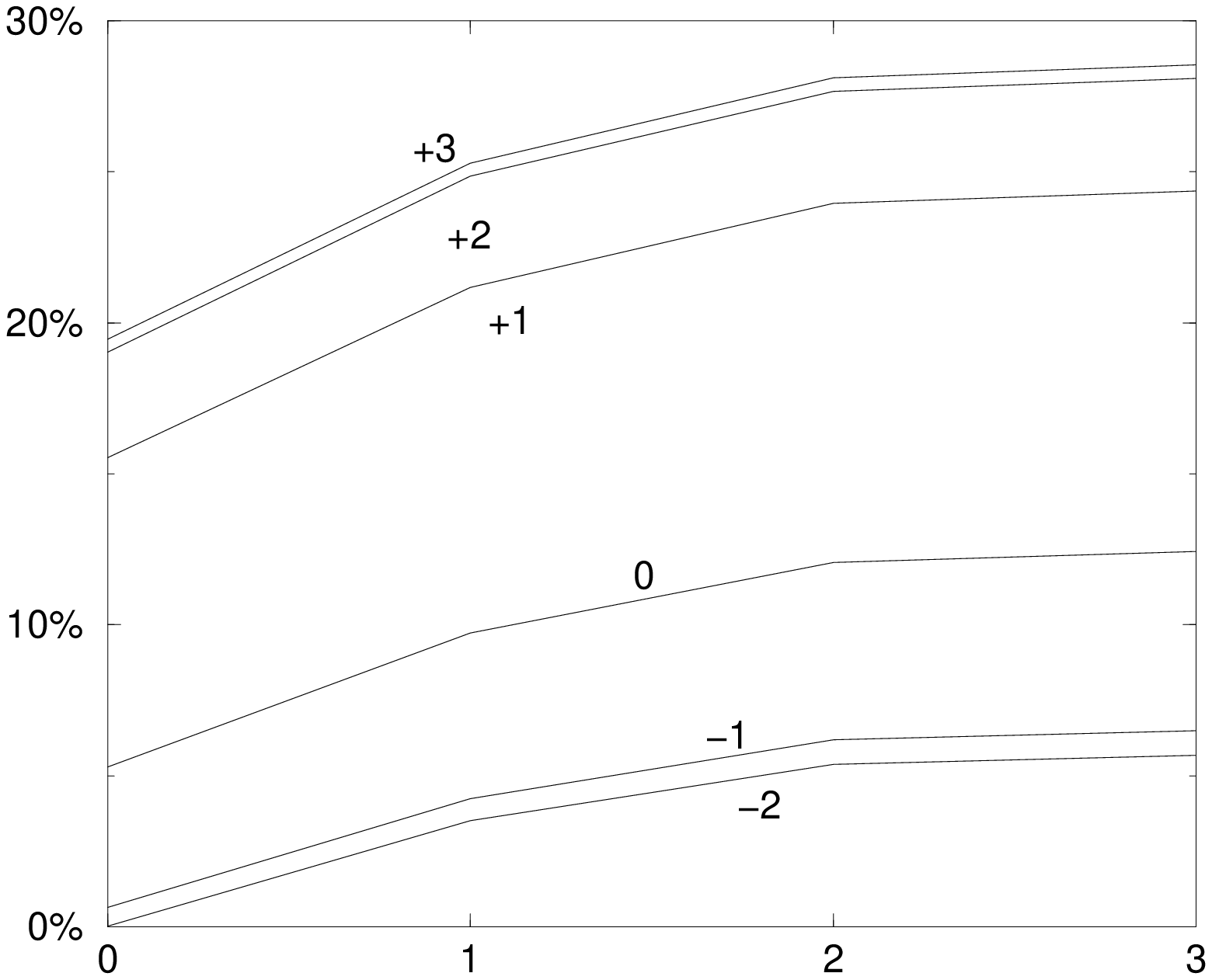,width=0.44\textwidth} 
    \\*[0cm] 
     & \parbox{0.5\textwidth}{\hspace{-11cm} $log_{10}(n)$}
      \end{tabular} 
\end{flushright} 
    \caption{Karlsruhe: The threshold reduction rate $\Gamma$ versus $log_{10}(n)$
for $e/R=0.1,q=1$ and different values of $s$. The labels correspond to
$log_{10}(s).$} 
\label{fig:karlmusigma}      
\end{figure}

In this section we vary the wall
permeability $n$ for $s=q=1$. 
We define again a threshold reduction rate by (\ref{gain}) in which $s$ is
replaced by $n$. The resulting reduction rate $\Gamma$ for Riga and Karlsruhe
are plotted respectively in Fig.\ref{fig:rigamu}  and Fig.\ref{fig:karlmu}
versus $log_{10}(n)$ for different values of $e/R$.\\
In the case of stationary solutions like for Karlsruhe, we find that $\Gamma$
is monotonically increasing versus $n$. We explain this increase by a change of
the geometry of the magnetic field lines in the fluid. When increasing $n$
the field lines in the fluid become perpendicular to the wall. As a result they
can close outside the fluid, decreasing the ohmic dissipation
in the fluid. As a result the total
dissipation decreases with $n$.\\
In the case of time-dependent solutions the
dissipation due to the eddy currents must be added to the previous total
dissipation. In that case,  
increasing the wall permeability still helps the magnetic field but prevents the electric currents
to close outside the fluid.
This can explain the difference of slope between the curves
(a) (negative slope) and (h) (positive slope) of Fig.\ref{fig:rigamu}.
Indeed in the case (a) the wall is probably larger than the skin depth and
the eddy currents dissipate more than the reduction of dissipation due  to
the change of geometry of the field lines. In case (h) the wall is so small
(smaller than the skin depth)
that the additional dissipation due to the eddy currents is negligible.\\
A common feature of Riga and Karlsruhe is that 
$\Gamma(s,n=q=1)=\Gamma(n,s=q=1)$ for $e/R\rightarrow \infty$. Such a relation
has already been found for the rigid body helical flow surrounded by a
conducting layer of infinite extent \cite{Mar95}.\\
For completeness we also calculated $\Gamma$ when both $s$ and $n$ are
changed (but still $q=1$). The corresponding curves are plotted in
Fig.\ref{fig:rigasigmamu} for Riga ($e/R=86\%$) and in
Fig.\ref{fig:karlmusigma} for Karlsruhe ($e/R=0.1$).\\   
\subsection{Influence of the fluid permeability}
\begin{figure}     \begin{flushright} 
  \begin{tabular}{@{\hspace{0cm}}l@{\hspace{0cm}}c@{\hspace{0cm}}} 
    \raisebox{4cm}{$\Lambda$} 
  \epsfig{file=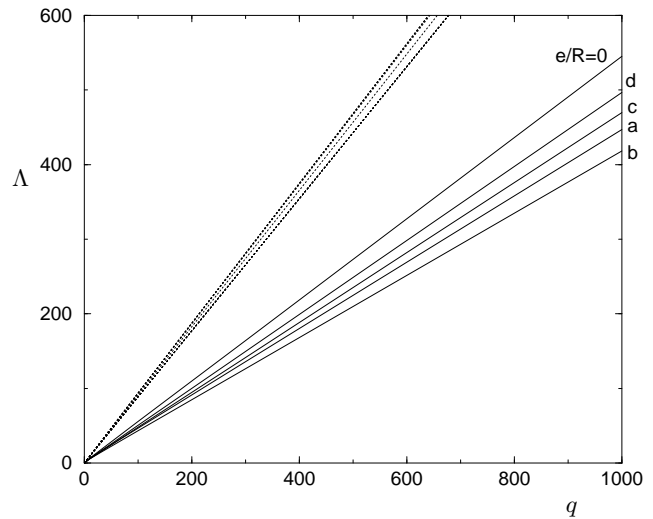,width=0.45\textwidth} 
    \\*[0cm] 
     & \parbox{0.5\textwidth}{\hspace{-11cm} $q$} 
    \end{tabular} 
\end{flushright} 
    \caption{The parameter $\Lambda$
versus $q$ for $n=s=1$ and different values of $e/R$.
The solid (dotted) curves correspond to Riga (Karlsruhe). The labels
correspond to those of Fig.\ref{fig:rigasigma}.}  
\label{fig:rigabille}      
\end{figure}

Here we look for the dynamo instability threshold assuming the use of a
ferro-fluid. The permeability of the wall is equal to the vacuum permeability. 
Therefore $s=n=1$ and $q=\mu_1/\mu_2$ is varied.
A simple way to estimate the benefit of using a ferro-fluid ($q>1$) is to
assume that the dynamo instability threshold does not vary significantly from
the case $q=1$. Then at the threshold $U(q)$ (resp. $\alpha_\perp(q)$) would behave like
$U(q=1)/q$ (resp. $\alpha_\perp(q=1)/q$).
Therefore the larger $q$ the smaller $U$ (or $\alpha_\perp$) would need to be, showing the possible
benefit of using a ferro-fluid.
However in this simple estimate the boundary conditions (\ref{BC})
in which the permeability jump between the fluid and the
surrounding wall is considered, is not satisfied.\\
When solving the problem with the full boundary conditions (\ref{BC})
 we find that in fact the threshold increases
with $q$. As a result using a ferro-fluid is less
interesting than suggested by the previous simple estimate. In
order to quantify how much less interesting, we calculate
$\Lambda=qR_m(q=1)/R_m(q)$ for Riga
and $\Lambda=qR_{\alpha}(q=1)/R_{\alpha}(q)$ for Karlsruhe versus $q$.
Then at the threshold $U(q)$
(resp. $\alpha_\perp(q)$) behaves like $U(q=1)/\Lambda$ (resp.
$\alpha_\perp(q=1)/\Lambda$).
The corresponding curves are plotted in Fig.\ref{fig:rigabille} with solid
(dotted) curves for Riga (Karlsruhe).  We find that $\Lambda$ is linear with $q$ and
that $1.8\le q/\Lambda\le2.4$ for Riga and $1.06\le q/\Lambda\le1.13$ for
Karlsruhe. Finally we conclude that using a ferro-fluid is still interesting
but again not as much as the simple previous estimate could give. Instead of
being equal to $q$ the gain on the flow intensity is about $q/2$ for Riga and
$q/1.1$ for Karlsruhe. 
\section{Conclusion}
For a dynamo laboratory experiment with stationary solutions like the
Karlsruhe experiment, the addition of an external wall with a conductivity $s$
larger than the fluid conductivity or with a permeability $n$ larger than
vacuum, leads to a reduction of the dynamo instability threshold. This
reduction is monotonous with $s$ and $n$. Typically the reduction can be as
high as 20$\%$ when only $s$ or $n$ is increased and up to 28$\%$ when both are
increased. This reduction is due to a change of geometry of the current lines
or the magnetic field lines leading to a reduction of the total ohmic
dissipation.\\  
For a dynamo laboratory experiment with non-stationary
solutions like the Riga experiment, the presence of some additional eddy
currents in the external wall reminiscent to a skin-effect changes the
previous results. In particular the reduction is not monotonous with $s$ nor
$n$. Indeed the eddy currents produce an additional dissipation which
can reduce the threshold drastically. As a result there is an optimum
conductivity $s$, permeability $n$ and wall thickness $e/R$ for which the
dynamo threshold is minimum. In Riga this optimum corresponds to a stagnant
layer of liquid sodium ($s=n=1$) of thickness $e/R=86\%$. Besides it is the
value actually used for the Riga experiment.\\
Finally the use of a ferro-fluid with a relative
permeability $q$ times larger than the vacuum permeability is interesting because the gain on the
velocity intensity or on the experiment dimension is about $q/2$ for Riga
and $q/1.1$ for Karlsruhe.
In practice
this could give some motivation for trying to obtain a ferro-fluid with a
permeability sufficiently large and homogeneous in space even in strong
motion.
\begin{acknowledgments}
R.A-Z. was supported by a Mexican Conacyt grant.
Part of the coding has been done by A.G. during a stay at the Laboratoire
des Ecoulements G\'{e}ophysiques et Industriels with a support from the
Institut National Polytechnique de Grenoble. \end{acknowledgments}
\bibliography{paper}

\begin{thebibliography}{15}
\expandafter\ifx\csname natexlab\endcsname\relax\def\natexlab#1{#1}\fi
\expandafter\ifx\csname bibnamefont\endcsname\relax
  \def\bibnamefont#1{#1}\fi
\expandafter\ifx\csname bibfnamefont\endcsname\relax
  \def\bibfnamefont#1{#1}\fi
\expandafter\ifx\csname citenamefont\endcsname\relax
  \def\citenamefont#1{#1}\fi
\expandafter\ifx\csname url\endcsname\relax
  \def\url#1{\texttt{#1}}\fi
\expandafter\ifx\csname urlprefix\endcsname\relax\def\urlprefix{URL }\fi
\providecommand{\bibinfo}[2]{#2}
\providecommand{\eprint}[2][]{\url{#2}}

\bibitem[{\citenamefont{Gailitis et~al.}(2000)\citenamefont{Gailitis,
  Lielausis, Dementiev, Platacis, Cifersons, Gerbeth, Gundrum, Stefani,
  Christen, H{\"a}nel et~al.}}]{Gai00}
\bibinfo{author}{\bibfnamefont{A.}~\bibnamefont{Gailitis}},
  \bibinfo{author}{\bibfnamefont{O.}~\bibnamefont{Lielausis}},
  \bibinfo{author}{\bibfnamefont{S.}~\bibnamefont{Dementiev}},
  \bibinfo{author}{\bibfnamefont{E.}~\bibnamefont{Platacis}},
  \bibinfo{author}{\bibfnamefont{A.}~\bibnamefont{Cifersons}},
  \bibinfo{author}{\bibfnamefont{G.}~\bibnamefont{Gerbeth}},
  \bibinfo{author}{\bibfnamefont{T.}~\bibnamefont{Gundrum}},
  \bibinfo{author}{\bibfnamefont{F.}~\bibnamefont{Stefani}},
  \bibinfo{author}{\bibfnamefont{M.}~\bibnamefont{Christen}},
  \bibinfo{author}{\bibfnamefont{H.}~\bibnamefont{H{\"a}nel}},
  \bibnamefont{et~al.}, \bibinfo{journal}{Phys.\ Rev. Lett.}
  \textbf{\bibinfo{volume}{84}}, \bibinfo{pages}{4365} (\bibinfo{year}{2000}).

\bibitem[{\citenamefont{Gailitis et~al.}(2001)\citenamefont{Gailitis,
  Lielausis, Platacis, Dementiev, Cifersons, Gerbeth, Gundrum, Stefani,
  Christen, and Will}}]{Gai01}
\bibinfo{author}{\bibfnamefont{A.}~\bibnamefont{Gailitis}},
  \bibinfo{author}{\bibfnamefont{O.}~\bibnamefont{Lielausis}},
  \bibinfo{author}{\bibfnamefont{E.}~\bibnamefont{Platacis}},
  \bibinfo{author}{\bibfnamefont{S.}~\bibnamefont{Dementiev}},
  \bibinfo{author}{\bibfnamefont{A.}~\bibnamefont{Cifersons}},
  \bibinfo{author}{\bibfnamefont{G.}~\bibnamefont{Gerbeth}},
  \bibinfo{author}{\bibfnamefont{T.}~\bibnamefont{Gundrum}},
  \bibinfo{author}{\bibfnamefont{F.}~\bibnamefont{Stefani}},
  \bibinfo{author}{\bibfnamefont{M.}~\bibnamefont{Christen}}, \bibnamefont{and}
  \bibinfo{author}{\bibfnamefont{G.}~\bibnamefont{Will}},
  \bibinfo{journal}{Phys.\ Rev. Lett.} \textbf{\bibinfo{volume}{86}},
  \bibinfo{pages}{3024} (\bibinfo{year}{2001}).

\bibitem[{\citenamefont{Stieglitz and M{\"u}ller}(2001)}]{Sti01}
\bibinfo{author}{\bibfnamefont{R.}~\bibnamefont{Stieglitz}} \bibnamefont{and}
  \bibinfo{author}{\bibfnamefont{U.}~\bibnamefont{M{\"u}ller}},
  \bibinfo{journal}{Phys. Fluids} \textbf{\bibinfo{volume}{13}},
  \bibinfo{pages}{561} (\bibinfo{year}{2001}).

\bibitem[{\citenamefont{Frick et~al.}(2002)\citenamefont{Frick, Khripchenko,
  Denisov, Pinton, and Sokoloff}}]{Fri02}
\bibinfo{author}{\bibfnamefont{P.}~\bibnamefont{Frick}},
  \bibinfo{author}{\bibfnamefont{S.}~\bibnamefont{Khripchenko}},
  \bibinfo{author}{\bibfnamefont{S.}~\bibnamefont{Denisov}},
  \bibinfo{author}{\bibfnamefont{J.-F.} \bibnamefont{Pinton}},
  \bibnamefont{and} \bibinfo{author}{\bibfnamefont{D.}~\bibnamefont{Sokoloff}},
  \bibinfo{journal}{Eur. Phys. J. B/Fluids} \textbf{\bibinfo{volume}{25}},
  \bibinfo{pages}{399} (\bibinfo{year}{2002}).

\bibitem[{\citenamefont{Gailitis et~al.}(2002)\citenamefont{Gailitis,
  Lielausis, E.Platacis, Gerbeth, and Stefani}}]{Gai02}
\bibinfo{author}{\bibfnamefont{A.}~\bibnamefont{Gailitis}},
  \bibinfo{author}{\bibfnamefont{O.}~\bibnamefont{Lielausis}},
  \bibinfo{author}{\bibnamefont{E.Platacis}},
  \bibinfo{author}{\bibfnamefont{G.}~\bibnamefont{Gerbeth}}, \bibnamefont{and}
  \bibinfo{author}{\bibfnamefont{F.}~\bibnamefont{Stefani}},
  \bibinfo{journal}{Reviews of Modern Physics} \textbf{\bibinfo{volume}{74}},
  \bibinfo{pages}{973} (\bibinfo{year}{2002}).

\bibitem[{\citenamefont{R{\"a}dler and Cebers}(2002)}]{Ceb02}
\bibinfo{editor}{\bibfnamefont{K.-H.} \bibnamefont{R{\"a}dler}}
  \bibnamefont{and} \bibinfo{editor}{\bibfnamefont{A.}~\bibnamefont{Cebers}},
  eds., \emph{\bibinfo{title}{Special issue on MHD Dynamo Experiments}}, vol.
  \bibinfo{volume}{28 (1-2)} (\bibinfo{publisher}{Magnetohydrodynamics},
  \bibinfo{year}{2002}).

\bibitem[{\citenamefont{Roberts}(1972)}]{Rob72}
\bibinfo{author}{\bibfnamefont{G.}~\bibnamefont{Roberts}},
  \bibinfo{journal}{Phil.\ Trans.\ R. Soc.\ Lond. A}
  \textbf{\bibinfo{volume}{271}}, \bibinfo{pages}{411} (\bibinfo{year}{1972}).

\bibitem[{\citenamefont{R{\"a}dler et~al.}(2001)\citenamefont{R{\"a}dler,
  Rheinhardt, Apstein, and Fuchs}}]{Rad01}
\bibinfo{author}{\bibfnamefont{K.}~\bibnamefont{R{\"a}dler}},
  \bibinfo{author}{\bibfnamefont{M.}~\bibnamefont{Rheinhardt}},
  \bibinfo{author}{\bibfnamefont{E.}~\bibnamefont{Apstein}}, \bibnamefont{and}
  \bibinfo{author}{\bibfnamefont{H.}~\bibnamefont{Fuchs}},
  \bibinfo{journal}{Nonlinear Processes in Geophysics}
  \textbf{\bibinfo{volume}{9}}, \bibinfo{pages}{171} (\bibinfo{year}{2001}).

\bibitem[{\citenamefont{Krause and R{\"a}dler}(1980)}]{Kra80}
\bibinfo{author}{\bibfnamefont{F.}~\bibnamefont{Krause}} \bibnamefont{and}
  \bibinfo{author}{\bibfnamefont{K.-H.} \bibnamefont{R{\"a}dler}},
  \emph{\bibinfo{title}{Mean--Field Magnetohydrodynamics and Dynamo Theory}}
  (\bibinfo{publisher}{Pergamon Press}, \bibinfo{year}{1980}).

\bibitem[{\citenamefont{Plunian and R{\"a}dler}(2002)}]{Plu02}
\bibinfo{author}{\bibfnamefont{F.}~\bibnamefont{Plunian}} \bibnamefont{and}
  \bibinfo{author}{\bibfnamefont{K.-H.} \bibnamefont{R{\"a}dler}},
  \bibinfo{journal}{Geophys. Astrophys. Fluid Dynamics}
  \textbf{\bibinfo{volume}{96}}, \bibinfo{pages}{115} (\bibinfo{year}{2002}).

\bibitem[{\citenamefont{A.Gailitis}(1990)}]{Gai90}
\bibinfo{author}{\bibnamefont{A.Gailitis}}, \emph{\bibinfo{title}{Topological
  Fluid Dynamics,edited by H.K. Moffatt and A. Tsinober}}
  (\bibinfo{publisher}{Cambridge University Press},
  \bibinfo{address}{Cambridge}, \bibinfo{year}{1990}).

\bibitem[{\citenamefont{Gailitis and Freiberg}(1980)}]{Gai80}
\bibinfo{author}{\bibfnamefont{A.}~\bibnamefont{Gailitis}} \bibnamefont{and}
  \bibinfo{author}{\bibfnamefont{Y.}~\bibnamefont{Freiberg}},
  \bibinfo{journal}{Magnetohydrodynamics} \textbf{\bibinfo{volume}{16}}
  (\bibinfo{year}{1980}).

\bibitem[{\citenamefont{Kaiser and Tilgner}(1999)}]{Kai99}
\bibinfo{author}{\bibfnamefont{R.}~\bibnamefont{Kaiser}} \bibnamefont{and}
  \bibinfo{author}{\bibfnamefont{A.}~\bibnamefont{Tilgner}},
  \bibinfo{journal}{Phys.\ Rev. E} \textbf{\bibinfo{volume}{60}},
  \bibinfo{pages}{2949} (\bibinfo{year}{1999}).

\bibitem[{\citenamefont{Sarson and Gubbins}(1996)}]{Sar96}
\bibinfo{author}{\bibfnamefont{G.}~\bibnamefont{Sarson}} \bibnamefont{and}
  \bibinfo{author}{\bibfnamefont{D.}~\bibnamefont{Gubbins}},
  \bibinfo{journal}{J. Fluid Mech.} \textbf{\bibinfo{volume}{306}},
  \bibinfo{pages}{223} (\bibinfo{year}{1996}).

\bibitem[{\citenamefont{Marty et~al.}(1995)\citenamefont{Marty, Ajakh, and
  Thess}}]{Mar95}
\bibinfo{author}{\bibfnamefont{P.}~\bibnamefont{Marty}},
  \bibinfo{author}{\bibfnamefont{A.}~\bibnamefont{Ajakh}}, \bibnamefont{and}
  \bibinfo{author}{\bibfnamefont{A.}~\bibnamefont{Thess}},
  \bibinfo{journal}{Magnetohydrodynamics} \textbf{\bibinfo{volume}{30}},
  \bibinfo{pages}{474} (\bibinfo{year}{1995}).

\end{thebibliography}
\end{document}